\PassOptionsToPackage{hypertexnames=false}{hyperref}
\documentclass[preprint,12pt]{elsarticle}
\usepackage{amssymb}
\usepackage{amsmath}
\usepackage{graphicx}
\usepackage{color}
\usepackage{pdfsync}
\usepackage{hyperref}
\usepackage[utf8]{inputenc}
\usepackage{relsize}

\newcommand{\etal}{{\it et al.}}
\newcommand{\bse}{\begin{subequations}}
\newcommand{\ese}{\end{subequations}}
\newcommand{\be}{\begin{equation}}
\newcommand{\ee}{\end{equation}}
\newcommand{\bea}{\begin{eqnarray}}
\newcommand{\eea}{\end{eqnarray}}
\newcommand{\kb}{k_{\mathrm{B}}}

\newcommand{\kernel}{\Gamma_{\!{\scriptscriptstyle \Omega}}}

\newcommand{\bet}{\Theta_{{\scriptscriptstyle \Omega}}}
\newcommand{\bethat}{\widehat{\Theta}_{{\scriptscriptstyle \Omega}}}
\newcommand{\omeg}{\widetilde{\omega}}

\newcommand{\vecphi}{\boldsymbol{\varphi}}

\journal{Nonlinear Science}
\begin{document}
\begin{frontmatter}
\title{Velocity modulus diffusion of self-propelled spherical and disk particles: A generalized Langevin approach}
\author{Pedro J. Colmenares}
\address{Universidad de Los Andes, Facultad de Ciencias, Departamento de Qu\'imica, M\'erida, Venezuela}
\ead{gochocol@gmail.com. Published in Nonlinear Science.}
\begin{abstract}

This research examines the average modulus of the velocity of an accelerated, self-propelled Brownian particle diffusing in a thermal fluid and confined by a harmonic external potential. The bath of harmonic oscillators at a constant temperature also interacts with the external field. The dynamics for both the sphere and a disk are partitioned into two distinct stochastic processes. The first describes the coarse-grained, time-dependent internal self-velocity generated by a set of independent Ornstein-Uhlenbeck processes, independent of the external field. This internal mechanism provides the initial velocity for the particle to diffuse within the fluid, which is modeled via a modified generalized Langevin equation as the second description. The system exhibits spontaneous fluctuations in the diffusive velocity due to an internal mechanism; however, these fluctuations decay over long times. Finally, the magnitude of the internal propulsion velocity in spherical coordinates is derived.
\end{abstract}

\end{frontmatter}
 
\section{Introduction}
\label{Sec1}

Research on self-propelled particles is currently a topic of intense interest. When the propulsion 
velocity remains constant, these are known as active Brownian particles (ABP). The ABP model 
serves as a fundamental prototype for describing diverse systems, such as bacterial colonies 
and motile cells. Comprehensive reviews by Romanczuk \etal \cite{Romanzuk}, Marchetti \etal 
\cite{Marchetti}, and Bechinger \etal \cite{Bechinger} document the theoretical progress 
in this field. Recent advancements have further expanded the theoretical toolkit, incorporating 
stochastic differential equations \cite{DepotZeno, Basu2D, Fodor, Kneu, Robertson, Caprini1, Caprini, 
Zhang1}, detailed analyses of statistical properties \cite{Martin}, dissipative dynamics \cite{Sinha}, 
and hydrodynamics \cite{RodriguezSolano}. Furthermore, the impact of large deviations in particle 
position on system dynamics has been extensively explored \cite{KetaEtal, Semeraro1EtAl, BuissonEtAl}. 
Such self-propelled behavior is ubiquitous, making it a crucial area of study even in the ecology of 
macroscopic populations \cite{Okubo}.

Generally, studies on ABP diffusion, such as those in Refs. \cite{Basu2D, Fodor, Caprini, Zhang1}, 
utilize an Ornstein-Uhlenbeck process (OUP) to model internal propulsion, followed by diffusion 
in the medium at a constant velocity. Martin \etal \cite{Martin} performed an analysis using a 
stochastic initial velocity. Additionally, fluid phase separation has been identified by several 
authors: Fily \etal \cite{FilyEtAl} examined short-range interactions without alignment between 
propulsion and diffusion velocities; Stenhammar \etal \cite{StenhammarEtAl} investigated the 
influence of particle geometry (disk vs. sphere); and Semeraro \etal \cite{Semeraro2EtAl} 
analyzed the process in terms of entropy production. Most approaches focus on determining 
position and velocity distributions. Notably, Gro{\ss}mann \etal \cite{GrobmannEtAl} described 
ABPs using an OUP self-propulsion with a piecewise velocity-dependent friction coefficient to 
maintain constant velocity. Das \etal \cite{Das2} employed an OUP framework without 
acceleration. Finally, Goswami {\it et al.} \cite{Goswami} model the diffusion of the ABP according to the overdamped Langevin equation with no inertia, with two noise sources, one from the thermal bath and a second from an OUP for self-propelling. They simulate run-and-tumble particles as bacteria using the Euler-Maruyama method to determine the mean square displacement of the ensemble average versus the time-averaged. Cherstvy {\it et al.} \cite{cherstvy} previously derived the dynamical equations for the same Langevin equation but only with thermal noise. 

Alternative investigations into ABPs focus on the role of an energy depot that provides internal speed. Romanczuk \etal \cite{Romanzuk} provide an excellent review of the core assumptions in this line of research. This mechanism involves a two-way energy transfer from the bath, internal dissipation, and conversion into kinetic energy proportional to the square of the velocity. These {\it ad hoc} factors are introduced to model ABP diffusion within a thermal reservoir, with velocity described by the overdamped Langevin equation (OLE). This approach was reformulated by Zeng \etal \cite{DepotZeno} using the multiplicative multi-noise Langevin equation derived by Denisov \etal \cite{Denisov}. They assigned independent noises to self-propulsion and diffusion, finding that the average speed depends on the cross-correlation intensity of the noises. While this approach can be analyzed via Hamilton's equations, it requires the energy depot's balancing equation (Eq. 48 in Ref. \cite{Romanzuk}). Incorporating this into the system Hamiltonian yields an equation of motion where the position is a functional derivative of the local squared velocity. Numerical analysis using the velocity expressions derived in this work renders a multiple-valued functional derivative. Consequently, the author suggests that the premises of the energy-depot approach should be reexamined for a consistent Hamiltonian description.

Beyond the referenced ABP models, other alternatives are suitable for analyzing self-propelled systems, particularly those where particles gradually accelerate due to specific internal mechanisms. Unlike standard ABP models, these particles randomly change their internal velocity and naturally reach a diffusive stationary state. 
Their velocity modulus (VM) can be obtained using more precise formulations, such as those derived by the author \cite{PJPRE5} or, more recently, by Netz \cite{Netz}; the internal process serves as the initial condition. This is the focus of the present research, hereafter referred to as the accelerated self-propelled diffusive particle (ASPDP) model.

The ASPDP model partitions the overall dynamics into two stochastic processes. The first describes the internal velocity driven by three OUPs, assuming the internal structure is insensitive to external perturbations and that the particle size remains consistent with the validity of the Langevin description. The second describes diffusion in a thermal bath of harmonic oscillators (HO) interacting with a harmonic external field, with initial conditions provided by the OUPs. This is implemented using a modified generalized Langevin equation (GLE)  derived in Ref. \cite{PJPRE5}. This framework is supported by its agreement with the molecular dynamics simulations of Daldrop \etal \cite{Daldrop} and its physical consistency with thermodynamic properties \cite{PJPRE6}. The resulting VM is analytically determined for spherical particles and extended to the planar case. Furthermore, the 3D-OUP is transformed into spherical coordinates to derive the stochastic differential equations for the VM and angular vector phase, a derivation that, to the author's knowledge, has not yet appeared in the literature.

The remainder of this paper is organized as follows: Section \ref{Sec2} derives the diffusive VM for spherical and disk-like particles; Section \ref{Sec3} analyzes the main results for both systems; and Section \ref{Sec4} provides concluding remarks. An appendix provides the derivation and numerical simulation of the velocity magnitude in spherical coordinates, including results for the disk in polar coordinates.

\section{Diffusion  velocity modulus}
\label{Sec2}

The velocity of a passive Brownian particle (PBP) diffusing in a thermal bath of harmonic oscillators under the influence of an external field was derived by Zwanzig \cite{Zwanzig2}. This result was subsequently shown to be exact for parabolic potentials by Glatzel and Schilling \cite{Glatzel} and Makri \cite{Makri}.

As previously mentioned, we employ the modified GLE derived in Ref. \cite{PJPRE5}. A compendium of its derivation is as follows. The system is composed of a tagged particle of mass $M$ immersed 
in a bath of $N$ harmonic oscillators held at temperature $T$. The whole system is subjected to the parabolic potential $V(q)=M\,\omega^{2}\, q^{2}/2$, where $q$ is the particle position. Early works on this topic are those from Lizy and Thotova \cite{lisy} and Thotova and Lizy \cite{TothovaLisy}. 

The  Hamiltonians for the particle $\mathcal{H}_{\mathrm{S}}$ with momentum $p$ and for the bath $\mathcal{H}_{\mathrm{B}}$ with degrees of freedom $\{p_{j},q_{j}\}$ are given as in the classical literature \cite{Zwanzig2, FordKacMazur}, with the difference that the latter includes the interaction with the field \cite{lisy, TothovaLisy}. They read,
\bea
\mathcal{H}_{\mathrm{S}}&=&\frac{p^{2}}{2\,M}+V(q),\\
\mathcal{H}_{\mathrm{B}}&=&\frac{1}{2}\sum_{j=1}^{N}\left[\frac{p_{j}^{2}}{m_{j}}+m_{j}\,\omega_{j}^{2}\left(q_{j}-\frac{\lambda_{j}\,q}{m_{j}\,\omega_{j}^{2}}\right)^{\!\!2}\right.+M\,\omega^{2}\,q_{j}^{2}\bigg],
\eea
where the last term in $\mathcal{H}_{\mathrm{B}}$ is the aforementioned field-bath interaction. $\lambda_{j}$ is the intensity of the bilinear coupling between the particle and each of the harmonic oscillators of the bath \cite {Zwanzig2}, having a mass $m_{j}$ and oscillating at frequency $\omega_{j}$.  Solving the equations for the bath's degrees of freedom and substituting them into the corresponding equation for the tag particle, as is usually done for the original GLE \cite{Hanggi2}, we get the desired modified GLE, which in vector form reads,
\bea
\dot{\mathbf{v}}_{\mathrm{d}}(t)&=&=-\Omega\,\mathbf{q}(t)-\int_{0}^{t}dy\,\mathbf{v}_{\mathrm{d}}(y)\,\kernel(\mid t-y\mid)+\mathbf{R}_{\scriptscriptstyle \Omega}(t),\label{MGLE}
\eea
where, unlike the original derivation, we use the subscript ``d'' to emphasize that the velocity is due only to diffusion.

The bath memory kernel $\kernel(t)$  is formally defined as
\bea
\kernel(t)&=&\frac{1}{M}\sum_{j=1}^{N}\frac{\lambda_{j}^{2}}{\alpha_{j}\,\beta_{j}}\,\cos\left(\alpha_{j}\,t\right),\label{EqTau}\\
&=&\frac{1}{M}\!\!\int_{-\sqrt{\kappa}\,\omega}^{\infty}d\omeg \sum_{j=1}^{N}\frac{\lambda_{j}^{2}}{\beta_{j}\,\omeg}\,\frac{\cos(\omeg\,t)}{\omeg}\,\delta(\omeg-\alpha_{j}),
\eea
where  $\alpha_{j}=(A_{j}/m_{j})^{1/2}$, $\beta_{j}=(A_{j}\,m_{j})^{1/2}$, $A_{j}= m_{j}\,\omega_{j}^{2}+M\,\omega^{2}$ and $\kappa$ represents the ratio of the tagged particle mass to that of a single bath particle. If the lower bound is chosen to be zero, the same result, $-\sqrt{\kappa}\,\omega$, is obtained. However, choosing the latter makes the kernel more physically significant than the former.
As explicitly mentioned by the author in Ref. \cite{PJPRE5}, “{\it this can be justified because the frequency of the oscillators is up-shifted by the amount of $\sqrt{\kappa}\,\omega$, as seen in the definition of $\alpha_{j}$. For the system to resonate in phase with the external field and obtain meaningful physical feedback, we must choose the negative value of the up-shift as the lower bound to increase the signal-to-noise ratio of the oscillator and achieve low interference in the response. Otherwise, if the lower bound is greater than or equal to zero, the overall effect is akin to the field and bath frequencies being out of phase, thereby generating an out-of-bounds function with no physical meaning. Additionally, the concept of negative frequency has a solid mathematical basis: it is defined by the phasor $e^{-i \,\omega\,t}$ rotating clockwise in the complex plane. Thus, according to Euler's formula, $\cos(\omega\, t)$ is equal to $(e^{i\omega t}+e^{-i\omega t})/2$, which is equivalent to two phasors rotating in opposite directions.}''

Defining the spectral density $J(\omeg)$ by \cite{Ingold,Pottier}
\be
J(\omeg)=\frac{\pi}{2}\sum_{j=1}^{N}\frac{\lambda_{j}^{2}}{m_{j}\,\omeg_{j}}\,\delta(\omeg-\omeg_{j}),
\label{Dirac}
\ee
then, the kernel in terms of the bath parameters reduces to:
\bea
\kernel(t)&=&\frac{2}{\pi\,M}\int_{-\sqrt{\kappa}\,\omega}^{\infty}d\omeg\,\frac{J(\omeg)}{\omeg}\cos(\omeg\,t),
\label{clas}
\eea
Next, we use Drude's spectral density for the coupling with the environment \cite{Ingold} 
\be
J(\omeg)=M\gamma_{0}\,\omeg\frac{\tau^{-2}}{\omeg^{2}+\tau^{-2}},
\label{Drude}
\ee
where $\gamma_{0}$ is the static value of the friction coefficient, and $\tau^{-1}$ is a cutoff frequency; that is, frequencies above the order of $\tau^{-1}$ are suppressed.
This choice leads to the bath parameters $\{m_{j},\omega_{j},\lambda_{j}\}$ not needing to be specified in the calculation of the kernel. Therefore, solving the integral gives a well-defined and analytic kernel, given by
\bea
\kernel(t)&=&\frac{\gamma_{0}}{\tau}\bigg\{e^{-t/\tau}-\frac{1}{\pi}\sinh\bigg(\frac{t}{\tau}\bigg)\bigg[\mathrm{Si}\bigg(H_{_{\!-}}\frac{t}{\tau}\bigg)+\mathrm{Si}\bigg(H_{_{\!+}}\frac{t}{\tau}\bigg)\bigg]\nonumber\\
&+&\frac{i}{\pi}\,\cosh\bigg(\frac{t}{\tau}\bigg)\bigg[\mathrm{Ci}\bigg(-i\,\frac{t}{\tau}\!\bigg)-\mathrm{Ci}\bigg(i\,\frac{t}{\tau}\bigg)\nonumber\\
&-&\mathrm{Ci}\bigg(H_{_{\!-}}\frac{t}{\tau}\bigg)+\mathrm{Ci}\bigg(H_{_{\!+}}\frac{t}{\tau}\bigg)\bigg]\bigg\} ,\label{sol1}\\
H_{_{\pm}}&=&\kappa\,\tau\,\omega\pm i,
\eea

The integral of the kernel over time gives its frequency-dependent friction coefficient 
$\gamma(\omega)$. In Ref. \cite{PJPRE5} it was shown that this result predicts a well-qualitative shape of the saturation curve obtained from the molecular dynamics simulation by Daldrop {\it et al.} \cite{Daldrop}.

The effective frequency $\Omega$ experienced by the PBP is not the nominal $\omega$ but 
\bea
\Omega&=&\omega^{2}\left[1+\frac{1}{M\,\omega^{2}}\sum_{j=1}^{N}\lambda_{j}^{2}\left(\frac{1}{m_{j}\,\omega_{j}^{2}}-\frac{1}{\alpha_{j}\,\beta_{j}}\right)\right]\!,\nonumber\\
&=&\omega^{2}\left(1+\sum_{j=1}^{N}\frac{\lambda_{j}^{2}}{A_{j}\,m_{j}\,\omega_{j}^{2}}\right),\nonumber\\
&=&\omega^{2}\,\bigg\{1-\frac{\gamma_{0}}{2\,\omega\,\bigg(\kappa\,\tau^{2}\,\omega^{2}-1\bigg)}\bigg[3\,\sqrt{\kappa}\nonumber\\
&-&2\,\kappa\,\tau\,\omega\bigg(1\!+\!\frac{2}{\pi}\arctan(\sqrt{\kappa}\,\tau\,\omega)\,\bigg)\bigg]\bigg\}\label{Omega},
\eea
where the last equation is obtained using Drude's spectral density as in the derivation of the analytic memory kernel.

 Finally, the modified generalized colored noise $\mathbf{R}_{\scriptscriptstyle \Omega}(t)$ keeps the functional form derived for the classical GLE with parameters modified by the field-bath interaction. It reads as:
  \be
\mathbf{R}_{\scriptscriptstyle \Omega}(t)=\frac{1}{M}\sum_{j=1}^{N}\lambda_{j}\left[\left(\mathbf{q}_{j}(0)-\frac{\lambda_{j}}{\beta_{j}\,\alpha_{j}}\,\mathbf{q}(0)\right)\cos(\alpha_{j}\,t)\right.+\frac{\mathbf{p}_{j}(0)}{\beta_{j}}\sin(\alpha_{j}\,t)\bigg],
\ee
where  $\{\mathbf{q}_{j}(0), \mathbf{p}_{j}(0)\}$ are the initial position and momentum of the bath oscillators.

The solution of the modified GLE is obtained by applying the Laplace transformation to Eq. (\ref{MGLE}). It renders:
\bse\bea
\mathbf{v}_{\mathrm{d}}(t)&=&\left<\mathbf{v}_{\mathrm{d}}(t)\right>_{\varphi}+\boldsymbol\varphi_{\mathrm{v}}(t),\label{Eqvt}\\
\left<\mathbf{v}_{\mathrm{d}}(t)\right>_{\varphi}&=&\mathbf{v}_{\mathrm{d}}(0)\,\chi(t) -\Omega\,\mathbf{q}(0)\int_{0}^{t}\!\!\!dy\,\chi(y),\label{vbar}\\
\chi(t)&=&\mathcal{L}^{-1}\left\{\frac{1}{k+\bethat(k)}\right\},\label{chivt}\\
\boldsymbol\varphi_{\mathrm{v}}(t)&=&\int_{0}^{t}ds\,\chi(t-s)\,\mathbf{R}_{\scriptscriptstyle \Omega}(s),\label{phinoise}
\eea\ese
where $\left<\cdots\right>_{{\varphi}}$ denotes an average over the effective colored noise $\boldsymbol\varphi_{\mathrm{v}}(t)$. The vectors $\mathbf{v}_{d}(0)$ and $\mathbf{q}(0)$ represent the initial diffusive velocity and the particle position, respectively. Furthermore, $\chi(t)$ is the system susceptibility to the external parabolic field, defined in terms of the inverse Laplace transform of $\bethat(k)$, where $\bet(t)=\kernel(t)+\Omega$. 
These equations can be extended to the ASPDP model by replacing the initial PBP values $\mathbf{v}_{\mathrm{d}}(0)$ and $\mathbf{q}(0)$ in Eq. (\ref{vbar}) with the equivalent propulsion vectors from the internal mechanism. These are identified with the subscript ``p'' for propulsion, namely $\mathbf{v}_{\mathrm{p}}(t)$ and $\mathbf{q}_{\mathrm{\,p}}(0)$, to emphasize that the diffusion evolves from the velocity and position of a set of OUPs describing the propulsion mechanism. Accordingly, the diffusive velocity magnitude in the ASPDP model is proposed as: 
\be
\mathbf{v}_{\mathrm{d}}(t)=\mathbf{v}_{\mathrm{p}}(t)\,\chi(t) -\Omega\,\mathbf{q}_{\mathrm{\,p}}(0)\int_{0}^{t}ds\,\chi(s)+\vecphi_{\mathrm{v}}(t),\label{vpj}
\ee
where the system susceptibility acts as a modulator of the time-dependent initial condition provided by the OUPs.
Each component of the vector $\mathbf{v}_{\mathrm{p}}(t)$ evolves according to:
\be
dv_{j}(t)=-\kappa_{j}\,v_{j}(t)\,dt + \epsilon_{j}\,dW_{j}(t);\hspace{0.2cm}j={x,y,z},\label{vpx}
\ee
where $W_{j}(t)$ represents an independent Wiener process, and $\{\kappa_{j},\epsilon_{j}\}$ denote the hydrodynamic drag and noise intensities for each component, respectively. The solution for this vector is well-established from stochastic calculus \cite{Gardiner1} as:
\bse\bea
\mathbf{v}_{\mathrm{p}}(t)&=&\mathlarger{\mathlarger{\sum}}_{j=1}^{3}\,\big(A(t,j)\!+\!B(t,j)\big)\,\widehat{k}_{j},\label{vpsol}\\
A(t,j)&=&v_{\mathrm{pj}}(0)\,e^{-\kappa_{j}t},\\
B(t,j)&=&\epsilon_{j}\int_{0}^{t}e^{-\kappa_{j}(t-s)}\,dW_{j}(s),
\eea\ese
where $\widehat{k}_{j}$ is the unit vector in the $j$-direction and $v_{\mathrm{pj}}(0)$ is the initial propulsion velocity component. We define $\left<\mathbf{v}_{\mathrm{d}}(t)\cdot\mathbf{v}_{\mathrm{d}}(t^{\prime})\right>_{\mathrm{w},\varphi}$ as the double average over the Wiener and colored noise distributions of the two-time correlation of Eq. (\ref{vpj}). Assuming the dynamics originate from a deterministic point—the center of the particle—with vanishing values for $\{ \mathbf{q}_{\mathrm{\,p}}(0),\mathbf{v}_{\mathrm{p}}(0)\}$, the correlation is given by:
\bea
\left<\mathbf{v}_{\mathrm{d}}(t)\cdot\mathbf{v}_{\mathrm{d}}(t^{\prime})\right>_{\mathrm{w},\varphi}&=&\chi(t)\,\chi(t^{\prime})\,\left<\mathbf{v}_{\mathrm{p}}(t)\cdot\mathbf{v}_{\mathrm{p}}(t^{\prime})\right>_{\mathrm{\!w}}+\int_{0}^{t}ds\,\chi(t-t^{\prime})\nonumber\\
&\times&\int_{0}^{t^{\prime}}ds^{\prime}\,\chi(t^{\prime}-s^{\prime})\left< \mathbf{R}_{\scriptscriptstyle \Omega}(s)\cdot \mathbf{R}_{\scriptscriptstyle \Omega}(s^{\prime})\right>_{\varphi}.\label{corvd}
\eea

Using Ito's calculus \cite{Gardiner1}, the two-time correlation $\left<\mathbf{v}_{p}(t)\cdot\mathbf{v}_{p}(t^{\prime})\right>_{\mathrm{w}}$ is determined from Eq. (\ref{vpsol}). For $t=t^{\prime}$, this reduces to:
\be
\big<\mathbf{v}_{p}(t)\cdot\mathbf{v}_{p}(t)\big>_{\mathrm{w}}=\mathlarger{\mathlarger{\mathlarger{\sum}}}_{j=1}^{3},\frac{\epsilon_{j}^{2}}{2,\kappa_{j}}\bigg(1-e^{-2,\kappa_{j} ,t}\bigg),\label{vmfinal1}
\ee
where the initial variance vanishes due to the deterministic initial condition.

Additionally, the noise correlation $\left<\boldsymbol\varphi_{\mathrm{v}}(t)\cdot\boldsymbol\varphi_{\mathrm{v}}(t^{\prime})\right>_{\varphi}$ given by Eq. (\ref{phinoise}), corresponds to the second term on the r.h.s. of Eq. (\ref{corvd}). Fox \cite{Fox} analytically solved this double integral by applying a double Laplace transform and invoking the fluctuation-dissipation theorem, which for the PBP model is expressed as \cite{PJPRE6}:
\be
\big< \mathbf{R}_{\scriptscriptstyle \Omega}(s)\cdot \mathbf{R}_{\scriptscriptstyle \Omega}(s^{\prime})\big>_{\varphi}=3\,\frac{\kb\,T}{M}\,\kernel(|s-s^{\prime}|),\label{FDTPJ1}
\ee
yielding the result:
\be
\big<\boldsymbol\varphi_{\mathrm{v}}(t)\cdot\boldsymbol\varphi_{\mathrm{v}}(t^{\prime})\big>_{\varphi}=3\,\frac{\kb\, T}{M}\,\bigg(\chi(|t-t^{\prime}|)-\chi(t)\,\chi(t^{\prime})\bigg).
\ee
Evaluated at equal times, and noting that $\chi(0)=1$ \cite{PJPRE6}, this simplifies to:
\be
\left<\boldsymbol\varphi_{\mathrm{v}}(t)\cdot\boldsymbol\varphi_{\mathrm{v}}(t)\right>_{\varphi}=3\,\frac{\kb\, T}{M}\,\bigg(1-\chi^{2}(t)\bigg).\label{vmfinal2}
\ee

Defining the diffusive VM as the root mean square of Eq. (\ref{corvd}) for $t=t^{\prime}$, denoted as $s_{\mathrm{d}}(t)$:
\be
s_{\mathrm{d}}(t)=\bigg[ \big<\mathbf{v}_{\mathrm{d}}(t)\cdot\mathbf{v}_{\mathrm{d}}(t)\big>_{\mathrm{w},\varphi}\bigg]^{1/2},
\ee
and combining Eqs. (\ref{vmfinal1}) and (\ref{vmfinal2}), we obtain:

\be
s_{\mathrm{d}}(t)=\Bigg[\chi^{2}(t)\,\mathlarger{\mathlarger{\mathlarger{\sum}}}_{j=1}^{3},\frac{\epsilon_{j}^{2}}{2\,\kappa_{j}}\bigg(1-e^{-2\,\kappa_{j}\,t}\bigg)+3\,\frac{\kb\,T}{M}\big(1-\chi^{2}(t)\big)\Bigg]^{1/2}.\label{VMfinal}
\ee

Equation (\ref{VMfinal}) is a general expression applicable to any 3D particle in the ASPDP model, where the external field does not affect the internal independent OUPs. The diffusion velocity in the medium follows physical principles similar to those of the PBP model: the system satisfies the fluctuation-dissipation theorem, and the susceptibility is a well-defined function of the external field frequency and system parameters. No specific alignment between diffusion and propulsion velocities is assumed, resulting in a constant VM at long times. Note that Eq. (\ref{VMfinal}) can also be derived directly from Eq. (\ref{vpj}) by considering the statistical independence of $\mathbf{v}_{\mathrm{p}}(t)$ and $\boldsymbol\varphi_{\mathrm{v}}(t)$ under the aforementioned initial conditions. Finally, this formulation can be extended to a two-dimensional description by omitting the $z$-component in Eq. (\ref{VMfinal}) and replacing the thermal term with $2\,\kb\,T\,\big(1-\chi^{2}(t)\big)/M$.

\section{Results and Discussion}
\label{Sec3}
Calculations were done for a set of parameters giving acceptable outcomes for the dynamical properties. It was discussed in Ref. \cite{PJPRE5} and subsequently extended to other properties in Ref. \cite{PJPRE6}.

\begin{figure}[ht]
\centering
\includegraphics[width=0.8\linewidth]{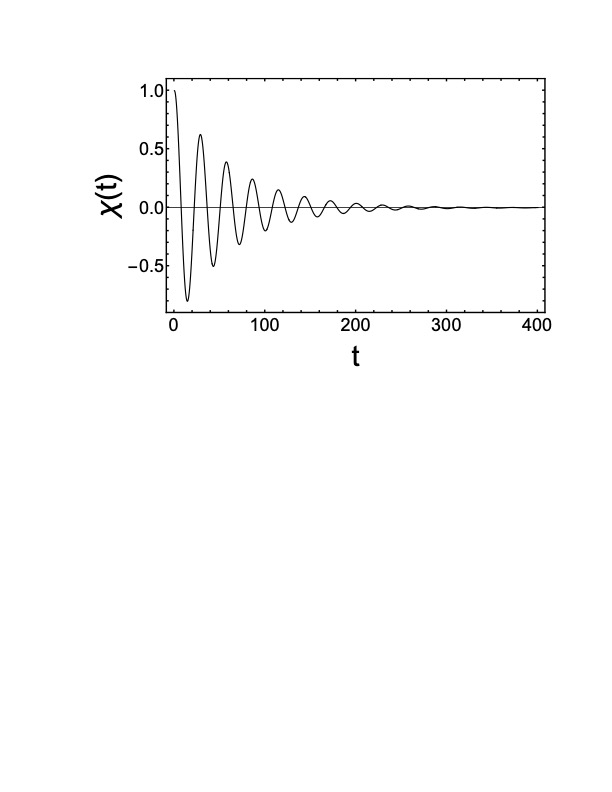}
\vspace{-7.5cm}
\caption{The susceptibility $\chi(t)$ defined by Eq. (\ref{chivt}), as previously calculated in Ref. \cite{PJPRE6}.}\label{Fig1}
\end{figure}

The calculation of the diffusive VM depends on the susceptibility $\chi(t)$ defined in Eq. (\ref{chivt}). This property was numerically evaluated in previous work, Fig (3) of Ref. \cite{PJPRE5} for $\kb=T= 1$ and $\{\omega, \kappa,\gamma_{0}\} = \{3, 2,1\}$. The procedure was to fit the kernel, Ec. (\ref{sol1}) by a train of three exponential decays. The fitting gives two distinct time scales: a fast one for diffusion and a slower one for the approach to the stationary state. The integration of the fitted kernel gives $\gamma(\omega)$, whose maximum equals 2, as the correct value for the chosen $\omega$. The susceptibility is obtained by inverting Eq. (\ref{chivt} for the already fitted memory kernel. It is shown in Fig. \ref{Fig1}.

\begin{figure}[ht]
\centering\includegraphics[width=0.8\linewidth]{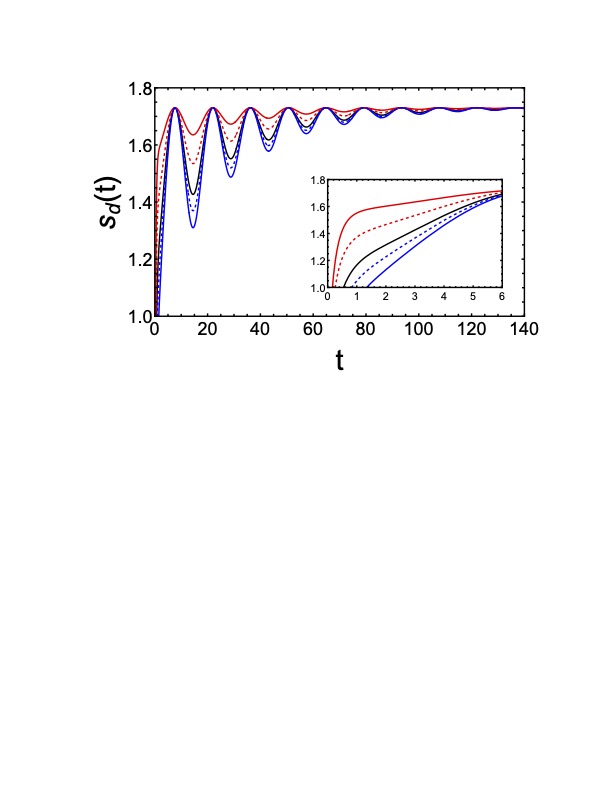}
\vspace{-7.5cm}
\caption{Diffusive VM for a spherical particle. The curves are identified by the parameter set $\{\epsilon, \kappa, \epsilon_{z}, \kappa_{z}\}$: $\{1,1,1,1\}$ (black), $\{0.5,0.5,1,1\}$ (blue), $\{1,1,0.5,0.5\}$ (dashed blue), $\{2,2,1,1\}$ (red), and $\{1,1,2,2\}$ (dashed red). The inset provides an amplification of the dynamics at short times.}
\label{Fig2}
\end{figure}

The diffusive VM for the sphere, as given by Eq. (\ref{VMfinal}), is displayed in Fig. \ref{Fig2}. For simplicity, we assume $\epsilon_{x}=\epsilon_{y}=\epsilon$ and $\kappa_{x}=\kappa_{y}=\kappa$. The figure illustrates results for various combinations of $\{\epsilon, \kappa, \epsilon_{z}, \kappa_{z}\}$, with the set $\{1, 1, 1, 1\}$ serving as a prototype (black curve).

The theory predicts the solid blue curve for the set $\{0.5, 0.5, 1, 1\}$, whereas a slight increase in the intensity of the minimum is observed when the parameter values are permuted to $\{1, 1, 0.5, 0.5\}$ (dashed blue curve). Nevertheless, the magnitude of these fluctuations decreases for higher intensities, as demonstrated by the solid red curve ($\{2, 2, 1, 1\}$) compared with the dashed red one ($\{1, 1, 2, 2\}$). The OUP modifies the particle acceleration, leading to a fluctuating VM that can be interpreted as changes in the velocity direction. These fluctuations decay over long times regardless of the parameter set, consistent with observations for the PBP \cite{PJPRE6}. As previously noted, the corresponding VM for a disk is given by:

\be
s_{\mathrm{d}}(t)=\bigg[\,\chi^{2}(t)\mathlarger{\mathlarger{\mathlarger{\sum}}}_{j=1}^{2}\,\frac{\epsilon_{j}^{2}}{2\,\kappa_{j}}\bigg(1-e^{-2\,\kappa_{j},t}\bigg) + 2\,\frac{\kb\,T}{M}\big(1-\chi^{2}(t)\big)\Bigg]^{1/2}.
\ee

\begin{figure}[ht]
\centering\includegraphics[width=0.7\linewidth]{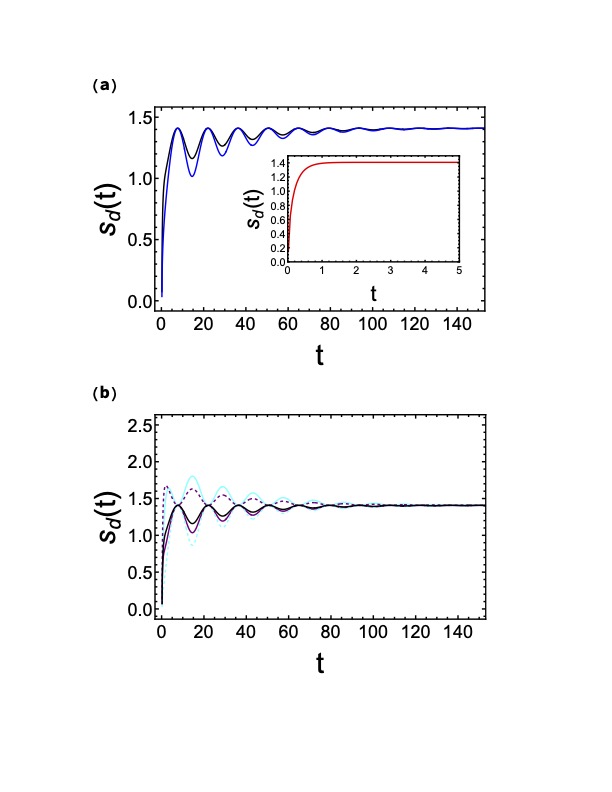}
\vspace{-2.2 cm}
\caption{Diffusive VM for the disk. Curves are identified by the set $\{\epsilon, \kappa\}$. Panel (a): results for $\{1,1\}$ (black), $\{0.5,0.5\}$ (blue), and $\{2,2\}$ (red), corresponding to the vanishing $z$-components of Fig. \ref{Fig2}. Panel (b): results for $\{1,1\}$ (black), $\{1,0.25\}$ (cyan), $\{0.25,1\}$ (dashed cyan), $\{1,1.75\}$ (purple), and $\{1.75,1\}$ (dashed purple).}
\label{Fig3}
\end{figure}

The results are presented in Fig. \ref{Fig3}. Panel (a) shows the diffusive VM for the disk by setting the $z$-direction intensities of the solid blue and red curves from Fig. \ref{Fig2} to zero; the original color markers are preserved for easy identification. Notably, the absence of the $z$-component decreases the VM while preserving the overall qualitative behavior. The effect of permuting parameter values is shown in panel (b) for the specific sets defined in the caption. Using $\{1, 1\}$ as a reference, we observe that swapping the parameter values (solid vs. dashed lines) inverts the curves

The fluctuating VM is a hallmark of non-equilibrium dynamics, where noise's disturbances eventually equilibrate with the particle's diffusion as the internal velocity reaches a steady state. This behavior was also observed for a single PBP \cite{PJPRE6} using the same parameters, albeit at longer timescales. In conclusion, the comparison between spherical and disk particles highlights how dimensionality and internal parameter anisotropy significantly alter the transient dynamics of self-propelled systems—a factor often overlooked in simplified 2D models. In summary, this proposal provides a rigorous stochastic framework for analyzing the transient and steady-state velocity of self-propelled particles, demonstrating that internal acceleration mechanisms lead to complex, geometry-dependent fluctuations before settling into a diffusive steady state.

\section{ Concluding remarks}
\label{Sec4}

This research determines the root-mean-square diffusive velocity magnitude of a Brownian particle in an external field, originating from rest with internal propulsion modeled by a set of independent OUPs. Diffusion within the thermal bath is described by a reformulation of the modified generalized Langevin equation derived in Ref. \cite{PJPRE5} for passive Brownian particles, utilizing the OUP velocities as initial conditions for the dynamics. Notably, the velocity magnitude depends solely on the model parameters without requiring additional ad hoc assumptions.

Both the spherical and disk systems exhibit spontaneous fluctuations in diffusive velocity due to a fluctuating acceleration generated by the internal OUP mechanism. However, these transient velocity fluctuations decay at long times, consistent with previous results for passive Brownian motion \cite{PJPRE6}. Such fluctuations arise from stochastic changes in the direction of the propulsion velocity. Consequently, the proposed model differs from standard active matter descriptions, with terminal velocity often assumed to be constant.

As a complement to this work, we provided the analytical derivation of the VM for a sphere. This result, which to our knowledge has not been reported in the literature, reduces to the established case for a disk. Simulations based on these equations predict discrepancies in the average velocity magnitude between the sphere and the disk at early stages; however, both systems exhibit saturation as time increases. There are other properties to determine, such as the mean square displacement \cite{Goswami, cherstvy}, to characterize the diffusion of the self-propelled BP. However, in the previous work \cite{PJPRE6} it was found that a BP (no propulsion) exhibits an ample domain of diffusivity ranging from sub to superdiffusion. Therefore, it is conceivable that such a set of behaviours in the diffusivity should appear in the ASPDP model.

The results shown above were conceived to investigate the same parameter domain used in the replication of the molecular simulations by Daldrop {\it et al.} in the physical outcomes of the ASPDP model. It is undeniable that real experimental parameters, as those used in Ref. \cite{Goswami}, should be used to test the descriptive power of the theory.

Although the modified GLE presented in this article is a linear differential equation and exact for parabolic potentials \cite{Glatzel, Makri}, the formalism can be extended to non-linear potentials with the caveat that the standard linear fluctuation-dissipation theorem breaks down. Rigorous derivations using projection-operator formalisms show that non-linear forces introduce position-dependent memory kernels, non-Gaussian noise, and cross correlations \cite{Jung}.

In summary, the key contributions of this study are twofold. First, the use of a generalized Langevin equation that explicitly accounts for the interaction of bath particles with the external field—an approach that replicates molecular dynamics simulations with high accuracy. Second, the integration of the Ornstein-Uhlenbeck process to describe the initial self-velocity. This framework could be further refined by employing alternative stochastic models for propulsion. Ultimately, this model serves as a foundation for describing nanomotors and other micro-swimmers that do not instantly reach terminal velocity but instead accelerate through an internal mechanism.

\section*{Acknowledgments}
No organization has funded this research. This work has profited from correspondence with Oscar Paredes-Altuve. The author thanks Nelson Pantoja for helpful discussions, Roy Little for proofreading the manuscript, and the anonymous reviewers for their valuable comments and suggestions.
\appendix
\section{Self-propelled OUP-VM in spherical coordinates}

Recalling that the components of the OUP-propelled particle velocities are described by, 
\be 
dv_{j}(t)=-\kappa_{j}\,v_{j}(t)\,dt + \epsilon_{j}\,dW_{j}(t);\hspace{0.2cm}j=\{x,y,z\},
\ee
we define $s_{\mathrm{p}}(t)$ as the magnitude of the vector $\mathbf{v}_{\mathrm{p}}(t)$, $\theta(t)$ as the polar angle along the $z$-axis, and $\phi(t)$ as the azimuthal angle in the $x\!-\!y$ plane. The Cartesian velocity components in spherical coordinates are then given by:
\bse
\bea
v_{x}(t)&=&s_{\mathrm{p}}(t)\,\sin\theta(t)\,\cos\phi(t),\label{ito1a}\\
v_{y}(t)&=&s_{\mathrm{p}}(t)\,\sin\theta(t)\,\sin\phi(t),\label{ito2a}\\
v_{z}(t)&=&s_{\mathrm{p}}(t)\,\cos\theta(t).\label{ito3a}
\eea
\ese

The derivation strategy is based on adapting Gardiner's polar coordinate method to a set of OUPs \cite{Gardiner1}. This involves products of stochastic variables in the Ito sense, which, with an appropriate definition of variables, can be transformed into the Stratonovich framework. Assuming implicit time dependency, let us define $\Phi=i\, v_{x}-v_{y}+e^{i\,\phi}\,v_{z}$, which, after substituting the velocity components, reduces to $\Phi=s_{\mathrm{p}}\,e^{i\,(\theta+\phi)}$. Therefore, defining $\mu=\ln s_{\mathrm{p}}$ as in the 2D case \cite{Gardiner1}, the 3D analog expression becomes:
\be
\ln \big[i\, v_{x}-v_{y}+e^{i\,\phi}\,v_{z}\big]=\mu+i\,(\theta+\phi),\label{A3}
\ee
which incorporates the $\phi$ angle dependency.

Expanding the differential of the previous equation up to the second order according to Ito's calculus yields:
\be
d\big[\mu+i\,(\theta+\phi)\big]=\frac{d(i\, v_{x}-v_{y}+e^{i\,\phi}\,v_{z})}{i\, v_{x}-v_{y}+e^{i\,\phi}\,v_{z}}-\frac{1}{2}\frac{\big[d(i\, v_{x}-v_{y}+e^{i\,\phi\,}\,v_{z})\big]^{2}}{(i\, v_{x}-v_{y}+e^{i\,\phi}\,v_{z})^{2}}.\label{A4}
\ee
Using the OUP components in the definition of $\Phi$ gives:
\bea
d(i\, v_{x}-v_{y}+e^{i\,\phi}\,v_{z})&=&-\big(i\,\kappa_{x}\, v_{x}-\kappa_{y}\,v_{y}+\kappa_{z}\,e^{i\,\phi}\,v_{z}\big)\,dt+\big(i\,\epsilon_{x}\, dW_{\!x}\nonumber\\
&-&\epsilon_{y}\,dW_{\!y}+\epsilon_{z}\,e^{i\,\phi}\,dW_{\!z}\big),\label{A5}
\eea

Substituting Eq. (\ref{A5}) into Eq. (\ref{A4}) and noting that the denominator is equal to $e^{-(\mu+i\,(\theta+\phi))}$, we obtain:
\bea
d\big[\mu+i\,(\theta+\phi)\big]&=&- e^{-\mu}e^{-i\, (\theta+\phi)}\bigg[\big(i\,\kappa_{x}\, v_{x}-\kappa_{y}\,v_{y}+\kappa_{z}\,e^{i\,\phi}\,v_{z}\big)\,dt\nonumber\\
&+&\big(i\,\epsilon_{x}\,dW_{\!x}-\epsilon_{y}\,dW_{\!y}+\epsilon_{z}\,e^{i\,\,\phi}\,dW_{\!z}\big)\bigg]-\frac{1}{2}\,e^{-2\,\mu}\,e^{-2\,i\,(\theta+\phi)}\nonumber\\
&\times&\big[d(i\, v_{x}-v_{y}+e^{i\,\phi}\,v_{z})\big]^{2}.\label{A6}
\eea

Solving for the product of the complex exponentials in the first and third terms \cite{Math}:
\bea
d\big[\mu+i\,(\,\theta+\phi)\big]&=&\bigg[A_{1}\!+\!\frac{1}{2}\,e^{-2\,\mu}\,C_{1}+i\,\bigg(B_{1}\!+\!\frac{1}{2}e^{-2\,\mu}\,D_{1}\bigg)\bigg]dt\nonumber\\
&+&e^{-\mu}e^{-i\,(\theta+\phi)}\bigg(i\,\epsilon_{x}\,dWx-\epsilon_{y}\,dW_{y}+\epsilon_{z}\,e^{i\,\phi}\,dW_{z}\bigg)\bigg],\label{A8}\\
A_{1}&=& -\sin\theta\bigg[\kappa_{x}\cos\phi\sin(\theta+\phi)-\kappa_{y}\sin\theta\cos(\theta+\phi)\bigg]\nonumber\\
&-&\kappa_{z}\cos^{2}\theta,\nonumber\\
C_{1}&=&(\epsilon_{x}^{2}-\epsilon_{y}^{2})\cos 2(\theta+\phi)-\epsilon_{z}^{2}\,\cos 2\theta,\nonumber\\
B_{1}&=&\epsilon_{x}\cos(\theta+\phi)\,dW_{x}-\epsilon_{y}\cos(\theta+\phi)\,dW_{y}+\epsilon_{z}\cos\theta\, dW_{z},\nonumber\\
D_{1}&=&\big(-\epsilon_{x}^{2}+\epsilon_{y}^{2}\big)\sin 2(\theta+\phi)+\epsilon_{z}^{2}\sin 2\theta.\nonumber
\eea

The real part of this equation can be rewritten as:
\bea
d\mu&=&\bigg(A_{1}+\frac{1}{2}e^{-2\,\mu}\,C_{1}\bigg)\,dt+\epsilon_{x}\sin(\theta+\phi)\,e^{-\mu}dW_{x}\nonumber\\
&-&\epsilon_{y}\,\cos(\theta+\phi)\,e^{-\mu}\,dW_{y}+\epsilon_{z}\,\cos\theta\,e^{-\mu}\,dW_{z}.
\label{A9}
\eea
The products $e^{\!-\mu}dW_{j}$ must be interpreted in the Stratonovich sense and transformed according to Ito rules. Following the transformations described in Ref. \cite{Gardiner1}:
\bse
\bea
\epsilon_{x}\,\sin(\theta+\phi)\,e^{-\mu}\,dW_{x}&=&-\frac{1}{2}\,\epsilon_{x}^{2}\,\sin^{2}(\theta+\phi) \,e^{-2\mu}\,dt\nonumber\\
&+&\epsilon_{x}\,\sin(\theta+\phi)\,dW_{x},\\
-\epsilon_{y}\,\cos(\theta+\phi)\,e^{\!-\mu}\,dW_{y}&=&-\frac{1}{2}\,\epsilon_{y}^{2}\,\cos^{2}(\theta+\phi)\, e^{\!-2\,\mu}\,dt\nonumber\\
&-&\epsilon_{y}\,\cos(\theta+\phi)\,dW_{y},\\
\epsilon_{z}\,\cos\theta\,e^{-\mu}\,dW_{z}&=&-\frac{1}{2}\,\epsilon_{z}^{2}\,\cos^{2}\theta \,e^{-2\mu}\,dt+\epsilon_{z}\,\cos\theta\, dW_{z}.
\eea
\ese

Substituting these into the original equation and noting that $d\mu=ds_{\mathrm{p}}/s_{\mathrm{p}}$ and $e^{-2\,\mu}=1/s_{\mathrm{p}}^{2}$, Eq. (\ref{A9}) becomes:
\bse
\bea
ds_{\mathrm{p}}&=&\bigg[A_{1}\,s_{\mathrm{p}}+\frac{1}{2\,s_{\mathrm{p}}}\bigg(C_{1}+E_{1}\bigg)\bigg]\,dt+dW_{1},\label{s1}\\
E_{1}&=&\epsilon_{x}^{2}\,\sin^{2}(\theta+\phi)+\epsilon_{y}^{2}\,\cos(\theta+\phi)+\epsilon_{z}^{2}\,\cos^{2}\theta,\nonumber\\
dW_{1}&=&\epsilon_{x}\sin(\theta+\phi)\,dW_{x}-\epsilon_{y}\cos(\theta+\phi)\,dW_{y}+\epsilon_{z}\cos\theta\,dW_{z}.\label{w1}
\eea
\ese

Similarly, the imaginary part of Eq. (\ref{A8}) yields:
\bse
\bea
d(\theta+\phi)&=&\bigg(B_{1}+\frac{1}{2\,s_{\mathrm{p}}^{2}}\,D_{1}\bigg)\,dt + \frac{dW_{2}}{s},\label{thetaphi1}\\
dW_{2}&=& \epsilon_{x}\cos(\theta+\phi)\,dW_{x}+\epsilon_{y}\sin(\theta+\phi)\,dW_{y}\nonumber\\
&-&\epsilon_{z}\sin\theta \,dW_{s}.
\eea
\ese

The system of Eqs. (\ref{s1}) and (\ref{thetaphi1}) is undetermined as there is no individual stochastic differential equation for $\theta$ or $\phi$. This can be resolved by defining $\widetilde{\Phi}=i\, v_{x}+v_{y}+e^{-i\,\phi}\,v_{z}$, which leads to $\ln\big[i\, v_{x}+v_{y}+e^{-i\,\phi\,}v_{z}\big]=\mu+i\,(\theta-\phi)$. Repeating the previous procedure, we get the set:
\bse
\bea
ds_{\mathrm{p}}&=&\bigg[A_{2}\,s_{\mathrm{p}}+\frac{1}{2\,s_{\mathrm{p}}}\bigg(C_{2}+E_{2}\bigg)\bigg]\,dt+dW_{3},\label{s2}\\
d(\theta-\phi)&=&\bigg(B_{2}+\frac{1}{2\,s_{\mathrm{p}}^{2}}\,D_{2}\bigg)\,dt + \frac{dW_{4}}{s},\label{thetaphi2}
\eea
\ese
where the following terms are defined:
\bse
\bea
A_{2}&=&-\sin\theta\bigg[\kappa_{x}\cos\phi\cos(\theta+\phi)-\kappa_{y}\sin\phi\sin(\theta+\phi)\nonumber\\
&-&\kappa_{z}\cos^{2}\theta\bigg],\\
C_{2}&=&(\epsilon_{x}^{2}-\epsilon_{y}^{2})\,\cos 2(\theta-\phi)-\epsilon_{z}^{2}\,\cos 2\theta,\\
E_{2}&=&\epsilon_{x}^{2}\,\sin^{2}(\theta-\phi)+\epsilon_{y}^{2}\,\cos^{2}(\theta\!-\!\phi)+\epsilon_{z}^{2}\,\cos^{2}\theta,\\
B_{2}&=&(\kappa_{x}-\kappa_{y})\,\sin\theta\,\cos(\theta-2\,\phi)\nonumber\\
&-&\frac{1}{2}\,(\kappa_{x}+\kappa_{y}-2\,\kappa_{z})\,\cos\phi,\\
D_{2}&=&(-\epsilon_{x}^{2}+\epsilon_{y}^{2})\,\sin2(\theta-\phi)+\epsilon_{z}^{2}\,\sin2\theta  ,\\
dW_{3}&=& \epsilon_{x}\sin(\theta-\phi)\,dW_{x}+\epsilon_{y}\cos(\theta-\phi)\,dW_{y}+\epsilon_{z}\cos\theta \,dW_{z},\\
dW_{4}&=& \epsilon_{x}\cos(\theta-\phi)\,dW_{x}-\epsilon_{y}\sin(\theta-\phi)\,dW_{y}-\epsilon_{z}\sin\theta \,dW_{z}. 
\eea
\ese

The final expression for $ds_{\mathrm{p}}$ is obtained by adding the two definitions from Eqs. (\ref{s1}) and (\ref{s2}). The equations for $d\theta$ and $d\phi$ are found by adding and subtracting Eqs. (\ref{thetaphi1}) and (\ref{thetaphi2}), respectively. These yield:
\bse
\bea
ds_{\mathrm{p}}&=&\bigg[\mathcal{P}\, s_{\mathrm{p}}+\frac{1}{2\,s_{\mathrm{p}}}\bigg(\mathcal{J}+\mathcal{C}\bigg)\bigg]\,dt +dW_{s},\label{seq}\\
d\theta&=&\bigg[\mathcal{K}+\frac{1}{2\,s_{\mathrm{p}}^{2}}\,\mathcal{L}\bigg]\,dt+\frac{1}{s_{\mathrm{p}}}dW_{\theta},\label{thetaeq}\\
d\phi&=&\bigg[\mathcal{M}+\frac{1}{2\,s_{\mathrm{p}}^{2}}\,\mathcal{N}\bigg]\,dt+\frac{1}{s_{\mathrm{p}}}dW_{\phi},\label{phieq}\\
dW_{\!s}&=&\epsilon_{x}\sin\theta\cos\phi\,dW_{\!x}+\epsilon_{y}\sin\theta\cos\phi\,dW_{\!y}+\epsilon_{z}\cos\theta\,dW_{\!z},\\
dW_{\!\theta}&=&\epsilon_{x}\cos\theta\cos\phi\,dW_{\!x}+\epsilon_{y}\cos\theta\sin\phi\,dW_{\!y}-\epsilon\sin\theta\,dW_{\!z},\\
dW_{\!\phi}&=&-\epsilon_{x}\sin\theta\sin\phi\,dW_{\!x}+\epsilon_{y}\sin\theta\cos\phi\, dW_{\!y},\eea
\ese
where $\mathcal{P}=(A_{1}+A_{2})/2$, $\mathcal{J}=(C_{1}+C_{2})/2$, $\mathcal{C}=(E_{1}+E_{2})/2$, $\mathcal{K}=(B_{1}+B_{2})/2$, $\mathcal{L}=(D_{1}+D_{2})/2$, $\mathcal{M}=(B_{1}-B_{2})/2$ and $\mathcal{N}=(D_{1}-D_{2})/2$. These are functions of the angles defined as:
\bse
\bea
\mathcal{P}&=&-\sin\theta\,\bigg(\kappa_{x}\,\cos\phi\,\sin(\theta+\phi)-\kappa_{y}\,\sin\phi\,cos(\theta+\phi)\bigg)\nonumber\\
& &-\kappa_{z}\,\cos^{2}\theta,\\
\mathcal{J}&=&\big(\epsilon_{z}^{2}-\epsilon_{y}^{2}\big)\, \cos2\,\phi-\epsilon_{z}^{2}\,\cos^{2}2\,\theta,\\
\mathcal{C}&=&\frac{1}{4}\,\bigg[\epsilon_{x}^{2}\,\bigg(2-\cos2(\theta-\phi)-\cos2(\theta+\phi)\bigg)+\epsilon_{y}^{2}\,\bigg(2+\cos2(\theta-\phi)\nonumber\\
&+&\cos2(\theta+\phi)\bigg)\bigg]+\frac{1}{2}\,\epsilon_{z}^{2}\,\bigg(1+\cos 2\theta\bigg),\\
\mathcal{K}&=&-\frac{1}{4}\,\bigg(\kappa_{x}+\kappa_{y}-2\,\kappa_{z}+\big(\kappa_{x}-\kappa_{y}\big)\,\cos 2\,\phi\bigg)\,\sin 2\,\theta,\\
\mathcal{L}&=&-\bigg[(\epsilon_{x}^{2}-\epsilon_{y}^{2}\big)\,\cos 2\,\phi-\epsilon_{z}^{2}\bigg]\,\sin 2\,\theta,\\
\mathcal{M}&=&\big(\kappa_{x}-\kappa_{y}\big)\,\cos\phi\,\sin^{3}\theta,\\
\mathcal{N}&=&-2\,\big(\epsilon_{x}^{2}-\epsilon_{y}^{2}\big)\,\cos 2\,\theta\,\sin 2\,\theta.
\eea
\ese

The system of Eqs. (\ref{seq})–(\ref{phieq}) constitutes the Ito version of the OUP in spherical coordinates. While analytic solutions are generally unavailable, they can be solved numerically by simulating each realization of $\{dW_{\!s}, dW_{\!\theta}, dW_{\!\phi}\}$ to obtain the propulsion VM $s_{\mathrm{p}}$.

Setting $\theta=\pi/2$ and $\kappa_{z}=\epsilon_{z}=0$, and assuming isotropic drag and noise intensity  $(\epsilon, \kappa)$ in the $x\!-\!y$ plane, the sphere equations reduce to the known disk equations in polar coordinates:
  \bse
  \bea
ds_{\mathrm{p}}&=&\left(-\kappa\,s_{\mathrm{p}}+\frac{\epsilon^{2}}{2\,s_{\mathrm{p}}}\right)\,dt\!+\!\epsilon\,dW_{\!s},\label{sdisk}\\
d\phi&=&\frac{\epsilon}{s_{\mathrm{p}}}\,dW_{\!\phi},\label{phidisk}\\
dW_{\!s}&=&\cos\phi\,dW_{\!x}+\sin\phi\,dW_{\!y},\label{wsdisk}\\
dW_{\!\phi}&=&-\sin \phi\,dW_{\!x}\!+\!\cos \phi\,dW_{\!y},\label{wphidisk}
\eea
\ese

These agree with Gardiner's OUP derivation in polar coordinates \cite{Gardiner1}. Notably, as Gardiner emphasizes, the noise definitions are valid only if the VM equation includes the stochastic correction term $\epsilon^{2}/(2s_{\mathrm{p}})$, a term omitted in Eq. (17) of Ref. \cite{Romanzuk}.

\begin{figure}[ht]
\centering
\includegraphics[width=0.6\linewidth]{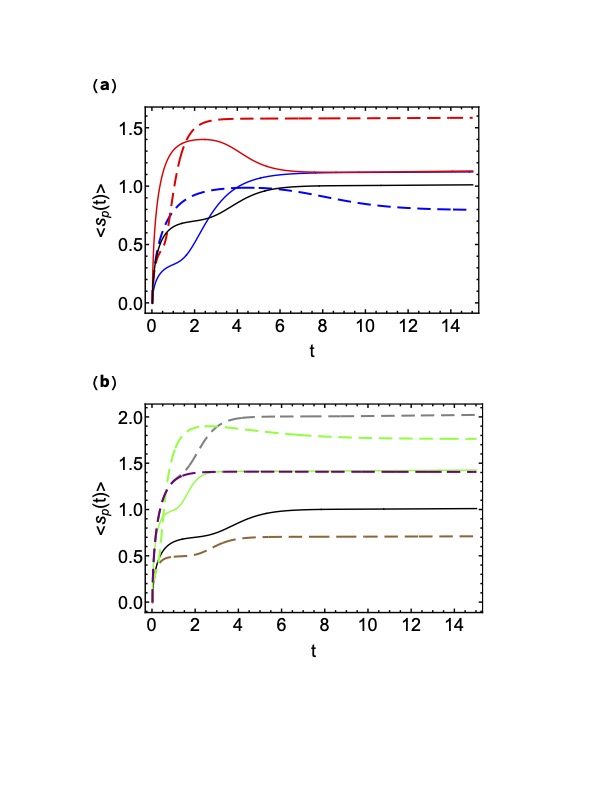}
\vspace{-2 cm}
\caption{Simulated averaged propelled VM of the sphere for various combinations of the set $\{\epsilon_{x},\epsilon_{y},\epsilon_{z}, \kappa_{x},\kappa_{y},\kappa_{z}\}$. (a) The curves correspond to those shown in Fig. \ref{Fig2}, with the same color identification, namely   $\{1,1,1,1,1,1\}$ (black) as reference, $\{1,1,2,1,1,2\}$ (dashed red), $\{2,2,1,2,2,1\}$ (red), $\{0.5,0.5,1,0.5,0.5,1\}$(blue), and $\{1,1,0.5,1,1,0.5\}$ (dashed blue). (b) This figure shows the results for other arbitrary parameter sets. The solid curves for $\{1,1,1,1,1,1\}$ (black) and $\{2,2,2,2,2,2\}$ (green) are shown as references; the remaining dashed curves are for $\{1,2,2,1,2,2\}$ (green), $\{2,2,2,1,1,1\}$ (gray), $\{2,1,1,2,1,1\}$ (purple), and $\{1,1,1,2,2,2\}$ (brown) .} 
\label{Fig4}
\end{figure}

The OU ordinary differential equations in spherical coordinates are solved \cite{Math} using continuous random functions associated with Wiener processes. These processes have a mean of zero and a standard deviation of $\sigma = 0.01$. This specific choice of $\sigma$ optimizes computational time while ensuring that the system's behavior remains close to its average trajectory, resulting in more stable and predictable simulations. For each time step, the stochastic velocity modulus and the two phases are added to their previous values. Consequently, the simulated average velocity modulus$\left<s_{\mathrm{p}}\right>_{\mathrm{\!w}}$   and the phases at the final time are calculated as their cumulative values divided by the number of realizations. It was found that for 2000 or more realizations, convergence is achieved for the modulus, though not for $\theta$ and $\phi$. Thus, it must be acknowledged that the results do not represent the exact solution of the ODEs, except for the convergent calculation of the velocity modulus. A similar outcome was observed for the disk.

\begin{figure}[ht]
\centering
\includegraphics[width=0.6\linewidth]{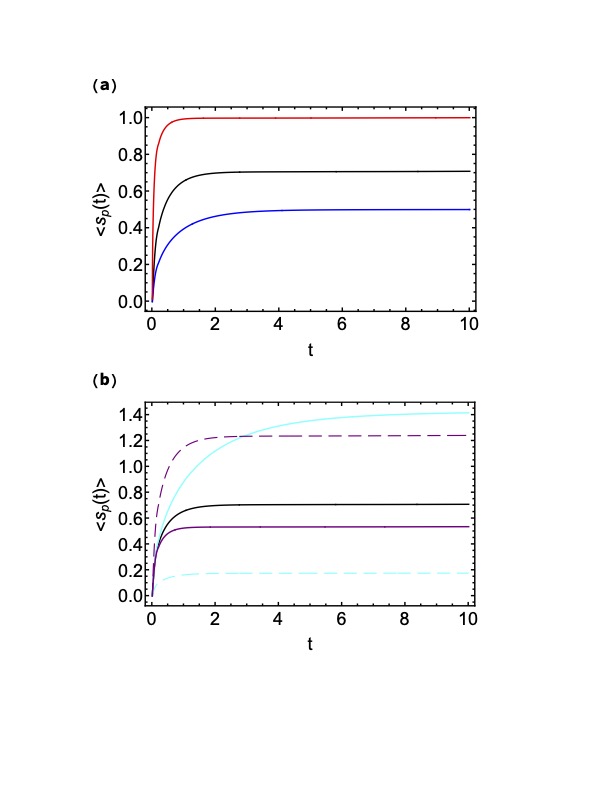}
\vspace{-2.1 cm}
\caption{Simulated averaged propelled VM of the disk for various combinations of the set $\{\epsilon,\kappa\}$. The graphs (a) and (b) correspond to the parameter sets of Fig. \ref{Fig3}a and \ref{Fig3}b, respectively, with the same color identification. }
\label{Fig5}
\end{figure}

Given the six parameters involved in simulating the average propelled 3D-VM, the parameter sets were selected to match those in Fig. \ref{Fig2} (using the same color identification), along with additional arbitrary sets to capture significant variations in the system's properties. These results are presented in Fig. \ref{Fig4}. The top panel corresponds to the $\{\epsilon_{x},\epsilon_{y},\epsilon_{z},\kappa_{x},\kappa_{y},\kappa_{z}\}$ sets from Fig. \ref{Fig2}: $\{0.5,0.5,1,0.5,0.5,1\}$ (solid blue), $\{1,1,0.5,1,1,0.5\}$ (dashed blue), $\{2,2,1,2,2,1\}$ (solid red), $\{1,1,2,1,1,2\}$ (dashed red), and $\{1,1,1,1,1,1\}$ (solid black) as a reference. The theory predicts early saturation for the dashed red curve, whereas the other sets saturate at later times—a physical requirement for diffusion dynamics to reach a steady speed in the long-time limit. The lower panel depicts predictions for other parameter sets, as detailed in the caption, where a perfect saturation (dashed purple) is observed. The initial "bumps" in the curves likely stem from the dominant effect of drag and noise intensities of the $x\!-\!y$ components over those of the polar coordinates; a greater disparity between these components appears to amplify the magnitude of the bump.

In contrast to the varied structural shapes obtained for the sphere, the results in polar coordinates consistently yield monotonic saturation curves, regardless of the $\{\epsilon,\kappa\}$ combination. These predictions are shown in the two panels of Fig. \ref{Fig5}, using the same parameter sets and color tags as in Fig. \ref{Fig3}. Ultimately, the structural richness observed for the sphere is lost for the disk, primarily due to the higher number of parameters involved in the former.


\end{document}